\documentclass[5p,times]{elsarticle}

\usepackage[T1]{fontenc}
\usepackage{cancel}
\usepackage{enumitem}
\usepackage{amsthm}
\usepackage{amsmath}
\usepackage{amssymb}
\usepackage{amsfonts}
\usepackage{graphicx}
\usepackage{dblfloatfix}
\usepackage{verbatim}
\usepackage{epstopdf}
\usepackage{tensor}
\usepackage{epsfig,multicol,bbm}
\usepackage{colortbl}
\usepackage{float}
\usepackage{multirow}
\usepackage{array}
\usepackage{ulem}
\usepackage[dvipsnames]{xcolor}
\usepackage{bm}
\usepackage{afterpage}
\usepackage{tabularx}
\usepackage{adjustbox}
\usepackage{mathrsfs}
\usepackage{tikz}
\usepackage{subcaption}
\usetikzlibrary{arrows.meta}
\usepackage{boxedminipage}
\usepackage{booktabs}
\usepackage{hyperref}
\usepackage{subcaption}
\newcolumntype{L}{>{\raggedright\arraybackslash}p{0.30\linewidth}}
\newcolumntype{Y}{>{\raggedright\arraybackslash}X}
\allowdisplaybreaks

\journal{Physics Letters B}

\begin{document}

\begin{frontmatter}

\title{Causal Horizons, Geodesic Completeness and Stability in Slow Contraction Cosmology}

\author[princeton]{Mariam Khaldieh}
\author[princeton]{Anna I. Rosenzweig}
\author[princeton]{Paul J. Steinhardt\corref{cor1}}
\ead{steinh@princeton.edu}
\cortext[cor1]{Corresponding author}

\address[princeton]{Department of Physics, Jadwin Hall, Princeton University, Princeton, NJ 08544, USA}

\begin{abstract}
We show that cosmological models with a semi-infinite phase of slow contraction (ekpyrosis) possess a combination of properties that can address several fundamental problems in cosmology, otherwise faced in contracting de Sitter phases or standard big bang expansion. In particular, flat or open slow contraction admits a stable past attractor that asymptotes to Minkowski space and is past geodesically complete, as well as a stable, flat, homogeneous, and isotropic future attractor with negligible Weyl curvature (and, therefore, negligible gravitational entropy). In bouncing cosmologies, this contracting attractor is terminated by a smooth, non-singular bounce that transforms the attractor properties at the end of contraction into the initial conditions for the subsequent expanding phase. Cosmologies incorporating a slow contraction phase have no particle horizon and therefore avoid the causal horizon problem. The past Minkowski attractor also generates an initial spectrum of vacuum-like quantum fluctuations on all wavelengths. Moreover, because the averaged expansion rate along past-directed geodesics is non-positive, models incorporating a semi-infinite phase of slow contraction also evade the Borde–Guth–Vilenkin theorem. By contrast, contracting de Sitter space possesses a finite particle horizon and becomes unstable in the presence of scalar fields, matter, or radiation.
\end{abstract}

\end{frontmatter}

\section{Introduction}
\label{sec:1}

Understanding the initial conditions of the universe remains a central problem
in cosmology. In addition to explaining the observed homogeneity, isotropy, and
flatness of the large-scale universe, a successful cosmology must also address issues of
causal structure, geodesic completeness, stability, and entropy.

In this paper, we consider cosmologies with a semi-infinite phase of slow contraction characterized by an equation of state $\varepsilon > 3$, where $\varepsilon\equiv (3/2)(1+ p/\varrho)$ for a phase with pressure $p$  and energy density $\varrho$~\cite{Khoury:2001wf}. A primary example is provided by bouncing cosmologies~\cite{IjjasSteinhardt2018}, in which the slow contraction phase is terminated by a smooth transition to expansion. Although some bouncing cosmologies are cyclical and so undergo finite periods of slow contraction, the models relevant here are single-bounce models in which the bounce is preceded by a semi-infinite phase of slow contraction.  We show how a slow contraction phase differs sharply from contracting de Sitter space and from singular Big Bang cosmology, exhibiting distinctive causal structure and global properties that are favorable for cosmological model building. Our analysis emphasizes the distinction between particle horizons and geodesic completeness, showing that slow contraction can be both past geodesically complete and free of a particle horizon, allowing arbitrarily large regions to share overlapping past light cones.

Beyond its causal structure, slow contraction exhibits several additional desirable properties. It admits dynamical attractor solutions, drawn from a large basin of initial conditions, running forward in time toward the bounce, that suppress spatial curvature, anisotropy, inhomogeneity, and Weyl curvature. Running backward in time, the same conclusion holds for a large basin of initial conditions in open universes, where spatial curvature dominates over any homogeneous anisotropy as the universe evolves toward the far past; in flat universes, the analogous Minkowski limit is reached only for the non-generic case of exactly homogeneous and isotropic initial data. This past attractor defines a natural vacuum state that seeds an initial spectrum of quantum fluctuations on all wavelengths often referred to as a Bunch-Davies vacuum~\cite{BirrellDavies1982}. Moreover, considering a past-directed timelike or null geodesic of a present-day observer integrated over the bounce and into the semi-infinite contracting phase, the averaged expansion rate along the geodesic is non-positive; consequently, models incorporating a semi-infinite phase of slow contraction evade the Borde--Guth--Vilenkin (BGV) theorem, consistent with past geodesic completeness in flat or open geometries. These features make slow contraction an attractive framework for constructing cosmological models consistent with observations.

Here we demonstrate that slow contraction exhibits the full combination of causal, geometric, dynamical, and semiclassical properties; see also  \cite{EricksonWesleySteinhardtTurok2004,Bari:2018,Cook:2020oaj,IjjasSteinhardt2021Entropy}.     The paper is organized as follows. Sections 2–4 analyze the causal structure of slow contraction and contrast it with contracting de Sitter phases, focusing on particle horizons. Section 5 studies geodesic completeness and the behavior of null and timelike geodesics.  Section 6 presents the conformal diagrams that illustrate the causal structure of  slow contraction cosmologies  when the expanding phase is radiation and matter-dominated or  when the expanding phase is $\Lambda$CDM and, hence, dominated by dark energy in the asymptotic future. Section 7 examines the effects of additional scalar fields and the stability of these backgrounds, highlighting the instability of contracting de Sitter space. Section 8 further clarifies the distinction between particle horizons and geodesic completeness. Section 9 shows that slow contraction evades the BGV theorem. Section 10 summarizes our conclusions.

\section{Background evolution during slow contraction}
\label{sec:2}

We consider a spatially flat Friedmann-Lemaitre-Robertson-Walker (FLRW) line element
\begin{equation}
    ds^2 = dt^2 - a^2(t)\, d\bm{x}^2,
\end{equation}
 where $a(t)$ is the scale factor and the co-moving time coordinate during the  contraction phase is chosen such that  $t<0$ and increasing.
Assuming that the energy density is dominated by a component with constant pressure-to-energy density ratio, $p/\varrho$, 
the contracting solution of the Friedmann equations for $t<0$ is: 
\begin{equation}
    a(t) = a_i\,(-t)^{1/\varepsilon},
    \label{eq:scalefactor}
\end{equation}
where $a_i>0$ is the value of $a(t)$ at some initial time $t_i$ . 
Slow contraction corresponds to   $\varepsilon>3$, {\it i.e.}, as $t \rightarrow 0^{-}$, the scale factor shrinks as a power law.
A bounce at some $t_b<0$ causes a transition to expansion before the universe would reach a crunch.

The contraction is ‘slow’ in the sense that the scale factor decreases less rapidly than  the Hubble radius,
\begin{equation}
 |H^{-1}| \propto |t| \propto a^{\varepsilon},  
\end{equation}
where $H \equiv \frac{\dot{a}}{a} = \frac{1}{\varepsilon t}$ is the Hubble parameter and dot represents the derivative with respect to $t$.
In typical scalar-field realizations, $\varepsilon$ can be very large, $\varepsilon \gg 3$.  The greater the value of $\varepsilon$, the slower the contraction. 
The weak  power-law behavior in Eq.~(\ref{eq:scalefactor})
 as $t\rightarrow -\infty$ plays a crucial role in determining the causal structure of the spacetime, as we will show.\\

\section{Comoving particle horizon in slow contraction}
\label{sec:3}

 We first consider the case
in which the slow contraction phase extends to the
infinite past, so that the comoving particle horizon is
\begin{equation}
    \chi_p(t_0) = \int_{-\infty}^{t_0} \frac{dt}{a(t)}.
\end{equation}

Substituting $a(t) = a_0\,(-t)^{1/\varepsilon}$
and dropping the constant prefactor, which does not affect the
convergence properties of the integral, we obtain
\begin{equation}
    \chi_p(t_0) \propto \int_{-\infty}^{t_0} \frac{dt}{(-t)^{1/\varepsilon}}
    = \int_{T_0}^{\infty} \frac{dT}{T^{1/\varepsilon}},
\end{equation}
where $T=-t$ so that $T>0$ for $t<0$ and $dt=-dT$. 
For $\varepsilon \ne 1$, the integral reduces to
\begin{equation}
    \chi_p(t_0) \propto
    \left.\frac{T^{1-\varepsilon^{-1}}}{1-\varepsilon^{-1}}\right|_{T_0}^{\infty}.
\end{equation}
When $\varepsilon> 1$, the exponent $1-\varepsilon^{-1}>0$, and $T^{1-\varepsilon^{-1}} \rightarrow \infty$ as
$T \rightarrow \infty$. Therefore the integral diverges:
\begin{equation}
    \chi_p(t_0) = \infty, \qquad \forall\, t_0<0,
\end{equation}
throughout the contracting phase. 
This result, also obtained in \cite{Bari:2018} for other types of bouncing cosmologies,  follows directly from the weak power-law dependence of the scale factor on $t$ when $\varepsilon>3$ at early times and is independent of the detailed microphysical realization of the slow-contraction phase.

An infinite comoving particle horizon implies that the comoving radius of
the past light cone of any observer grows without bound as $t \rightarrow -\infty$.  Consequently, there is no particle horizon: any two comoving points with finite comoving separation 
    have overlapping past light cones at sufficiently early times.

\medskip

\noindent
{\bf Remark 1.}
The absence of a particle horizon established here depends only on the
behavior of the conformal-time integral $\int dt/a(t)$ and does not imply past geodesic completeness of the spacetime.
Issues of geodesic completeness, and their dependence on spatial curvature
and global properties of the solution, are addressed separately in
Section~\ref{sec:8}.

\medskip

\noindent
{\bf Remark 2.}
In expanding de Sitter space, the cosmological event horizon is associated with marginally trapped surfaces. 
By contrast, in past-complete slow contraction the conformal-time integral
diverges and the particle horizon is absent; the  marginal radius
$R\sim |H|^{-1}$ recedes to infinity as $t\rightarrow -\infty$, and the geometry
approaches Minkowski space.
Consequently, there is no natural cosmological screen extending into the
a\-symptotic past.
See discussion of conformal diagrams in Section~\ref{sec:6}.

\section{Contracting de Sitter in flat slicing: a finite past horizon}
\label{sec:4}

We contrast the case of slow contraction with a contracting de Sitter spacetime. Throughout our discussion of de Sitter (this section and Section~\ref{sec:5}), we denote by $H_\Lambda \equiv \sqrt{\Lambda/3} > 0$ the de Sitter scale, a positive constant set by the cosmological constant $\Lambda$. Consider a contracting de Sitter phase with scale factor
\begin{equation}
    a_{\text{dS}}(t) = e^{-H_\Lambda t},
\end{equation}
where $ H_\Lambda>0$ for $t<0$. The universe contracts from infinite size at $t\rightarrow -\infty$ to a finite size at $t\rightarrow 0^{-}$. This description corresponds to the flat slicing of de Sitter space.

The comoving particle horizon at time $t_0<0$ is
\begin{equation}
    \chi_p^{(\text{dS})}(t_0) = \int_{-\infty}^{t_0} \frac{dt}{a_{\text{dS}}(t)} = \frac{1}{H_\Lambda}\,e^{H_\Lambda\,t_0},
\end{equation}
which is finite for every finite $t_0$. The corresponding physical particle horizon size is constant:
\begin{equation}
    d_p^{(\text{dS})}(t_0) = a_{\text{dS}}(t_0)\,\chi_p^{(\text{dS})}(t_0) = \frac{1}{H_\Lambda}.
\end{equation}

Thus, even though the contracting de Sitter phase extends to $t=-\infty$ and each spatial slice is infinite, the comoving region in causal contact with a given observer is bounded; the resulting particle horizon has  physical radius {$H_\Lambda^{-1}$}.  Not all comoving pairs of points are guaranteed to have been in causal contact, even if the contracting de Sitter phase extends to $t=-\infty$. This finite horizon is associated with the presence of marginally trapped surfaces in the contracting geometry.

The difference with slow contraction stems from the different asymptotic behavior of the scale factor as $t \rightarrow -\infty$:
\begin{itemize}
    \item In rapidly contracting de Sitter, $a(t)$ grows exponentially as $t \rightarrow -\infty$, while the integrand $1/a(t)$ decays exponentially, so the integral $\int dt/a(t)$ converges. The de Sitter scale $H_{\Lambda}$ is constant.
    \item In {\it slow} contraction with $\varepsilon>3$, the scale factor decays as a power law, $a(t)\propto(-t)^{1/\varepsilon}$, and the integral $\int dt/a(t)$ diverges. The Hubble radius $|H^{-1}|$ shrinks rapidly relative to the scale factor.
\end{itemize}

\section{The causal structure of global de Sitter}
\label{sec:5}

In Section~\ref{sec:4}, we compared slow contraction to contracting 
de~Sitter in spatially flat slicing. However, the causal structure takes on a 
different appearance when de~Sitter is expressed in global (closed)
slicing. Since the closed slicing contains a ``neck,'' {\it i.e.}, a moment of minimal 
spatial volume, it is natural to wonder whether this implies that the entire 
spatial slice must lie within a single causal patch. We now show that this is 
not the case.\footnote{Note that different slicings of de~Sitter correspond to distinct families of comoving observers that are not at rest with respect to one another. Consequently, observers may disagree about expansion versus contraction and about the presence or absence of a particle horizon, even though the underlying causal structure of the spacetime is the same.}

In closed slicing, the de~Sitter line element may be written as
\begin{equation}
    ds^2 =  dt^2 - a^2(t)\, d\Omega_3^2,
\end{equation}
with
\begin{equation}
   a(t) = \frac{1}{H_\Lambda} \cosh(H_\Lambda t).
\end{equation}
The spatial hypersurfaces are 3-spheres of radius $a(t)$ whose minimum,
\begin{equation}
    a_{\mathrm{min}} = \frac{1}{H_\Lambda},
\end{equation}
is reached at $t=0$ and is commonly referred to as the ``neck'' of de~Sitter. One might 
suspect that because the spatial radius is precisely {$H_\Lambda^{-1}$}, any two points on 
this hypersurface must have been in causal contact prior to $t=0$. This is not the case, though.

To see why, consider the comoving radial coordinate $\chi$ on $S^3$, where
$0 \le \chi \le \pi$. Light propagates as
\begin{equation}
    d\chi = \frac{dt}{a(t)}.
\end{equation}
The maximum comoving distance a null ray could have traversed from 
$t=-\infty$ to the neck at $t=0$ is
\begin{align}
    \Delta \chi_{\max}
    &= \int_{-\infty}^{0} \frac{dt}{a(t)} 
      = H_\Lambda \int_{-\infty}^{0} \frac{dt}{\cosh(H_\Lambda t)} 
      \\
    &= \arctan\big(\sinh(H_\Lambda t)\big)\,\Big|_{t=-\infty}^{0} 
      = \frac{\pi}{2}.
\end{align}
By contrast, the maximum 
geodesic separation between two antipodal points is twice as large, $\chi_{\mathrm{max}} = \pi$.

Thus, even with infinite amount of past proper time, null rays can explore at
most half the comoving diameter of the spatial $S^3$ at $t=0$. 
The remaining portion of the 3-sphere is outside the observer's causal past. 

This illustrates a key contrast with slow contraction. In closed de~Sitter, the 
exponential behavior of the scale factor in the far past causes the integral 
$\int dt/a(t)$ to converge, producing a finite past horizon. The spatial volume 
at the neck does not determine the causal structure; rather, the limiting 
behavior of $a(t)$ as $t \rightarrow -\infty$ does. In slow contraction, by contrast, 
the power-law behavior $a(t) \propto (-t)^{1/\varepsilon}$  with $\varepsilon >3$ causes the integral to 
diverge, eliminating the past horizon entirely.

\section{Conformal diagrams for slow contraction cosmology}
\label{sec:6}

We now consider a universe that undergoes a $\Lambda$CDM expansion phase which is preceded by a history that includes a  slow contraction phase and asymptotes a  Minkowski spacetime in the  infinite past. A universe where late-time expansion is  driven by $\Lambda$ and one where it is driven by a time-varying (quintessence) dark energy component that dominates in the asymptotic future can be described by  causal diagrams with an equivalent global structure.

The global spacetime does not encounter any spacelike singularity in the past, as the slow contraction phase is terminated by a classical (non-singular) bounce occurring on the $\bar{\chi}$ hypersurface of constant $\bar{\eta}$ as indicated in Fig.~2. 
We use barred conformal coordinates $(\bar{\eta}, \bar{\chi} )$ to define the conformal metric 
\begin{equation}
    ds^2 = a^2(\bar{\eta})\big(d \bar{\eta}^2 -d\bar{\chi}^2-\bar{\chi}^2d\bar{\Omega}^2 \big)
\end{equation}
and run in the semi-infinite range $0>\bar{\eta}>-\infty$ and $+\infty>\bar{\chi}>0$. To describe each phase of cosmic evolution, we track the appropriate scale factor corresponding to the given phase. We match the cosmological solutions on the boundaries, such that each phase smoothly transitions to the following phase.  

The conformal diagram is  drawn in unbarred coordinates $(\eta, \chi)$ using the compactification scheme given in \cite{Mukhanov2005}, namely
\begin{equation}
    \bar{\eta}= \frac{\sin(\eta)}{\cos(\eta)+\cos(\chi)}, \hspace{1cm}   \bar{\chi}= \frac{\sin(\chi)}{\cos(\eta)+\cos(\chi)}.
\end{equation}
The interior curves are spatial hypersurfaces of constant
cosmological conformal time $\bar\eta$ and worldlines of comoving observers
at constant $\bar\chi$ during each phase. The unbarred coordinates extend
to cover the global de Sitter space. The flat-slicing coordinates
$(\bar\eta, \bar\chi)$ parameterize only a portion of it.
The radial null rays satisfy
\begin{equation}
   d\chi=\pm d\eta.
\end{equation} 

Note that the  area assigned to each phase does not accurately represent its relative duration in the given cosmological model in each diagram.

\paragraph{Horizons} As shown in Fig.~1,  the conformal diagram of a universe with an asymptotic past Minkowski, slow contraction, bounce and radiation-matter dominated expansion phase reveals the absence of a particle horizon, as we found earlier in Sec.~\ref{sec:3}, and the diagram is bounded by a null-like hypersurface $\mathcal{J^-}$. An event horizon is absent here, as well. In Fig.~2, the diagram corresponds to a universe where matter domination is followed by late-time acceleration.  As the cosmological constant comes to dominate, an event horizon is formed and the upper part of Fig.~1 resembles the usual de Sitter spacetime in flat coordinates. In fact, the causal structure of such a universe globally resembles flat de Sitter, but the details pertaining to the evolution of the scale factor $a(\eta)$  and 
the Hubble radius $|H(\eta)|^{-1}$
 differ significantly.

\paragraph{The bounce}  A non-singular bounce occurs on a constant time hypersurface at some time before the universe would reach the putative crunch (at $\eta<0$). The exact features of this hypersurface in the diagram depend on the physics of the bounce.

\begin{figure}[!htbp]
    \centering

    \begin{subfigure}{\columnwidth}
        \centering
        \includegraphics[width=1.2\columnwidth]{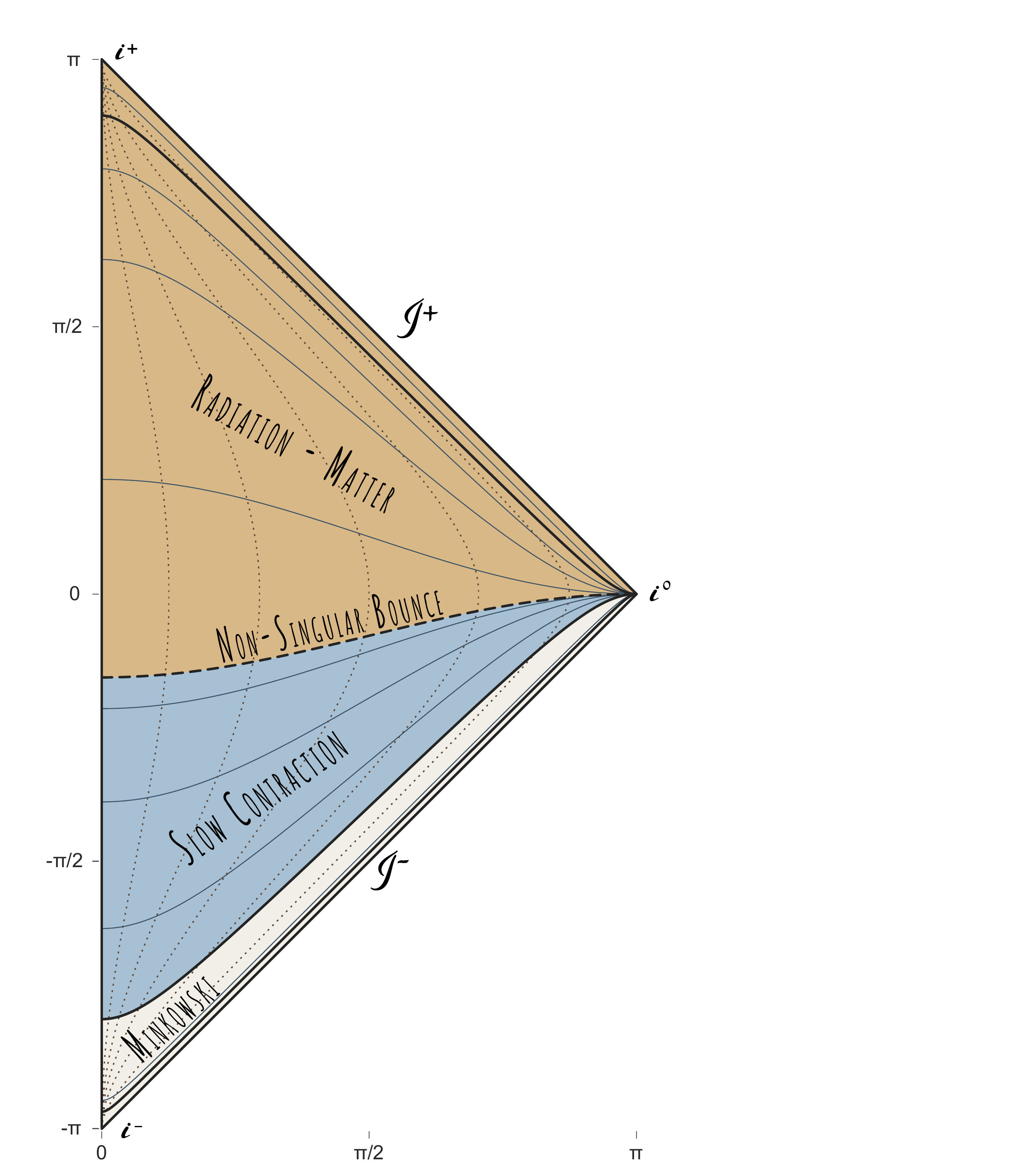}
        \caption{Figure 1.  Conformal Diagram in $\eta-\chi$ plane of a universe with an asymptotic past Minkowski, a slow contraction phase ending with a non singular bounce (dashed curve) and smoothly transitioning into a radiation-matter dominated phase in the asymptotic future. }
        \label{fig:first}
    \end{subfigure}

    \vspace{0.5em}

    \begin{subfigure}{\columnwidth}
        \centering
        \includegraphics[width=1.2\columnwidth]{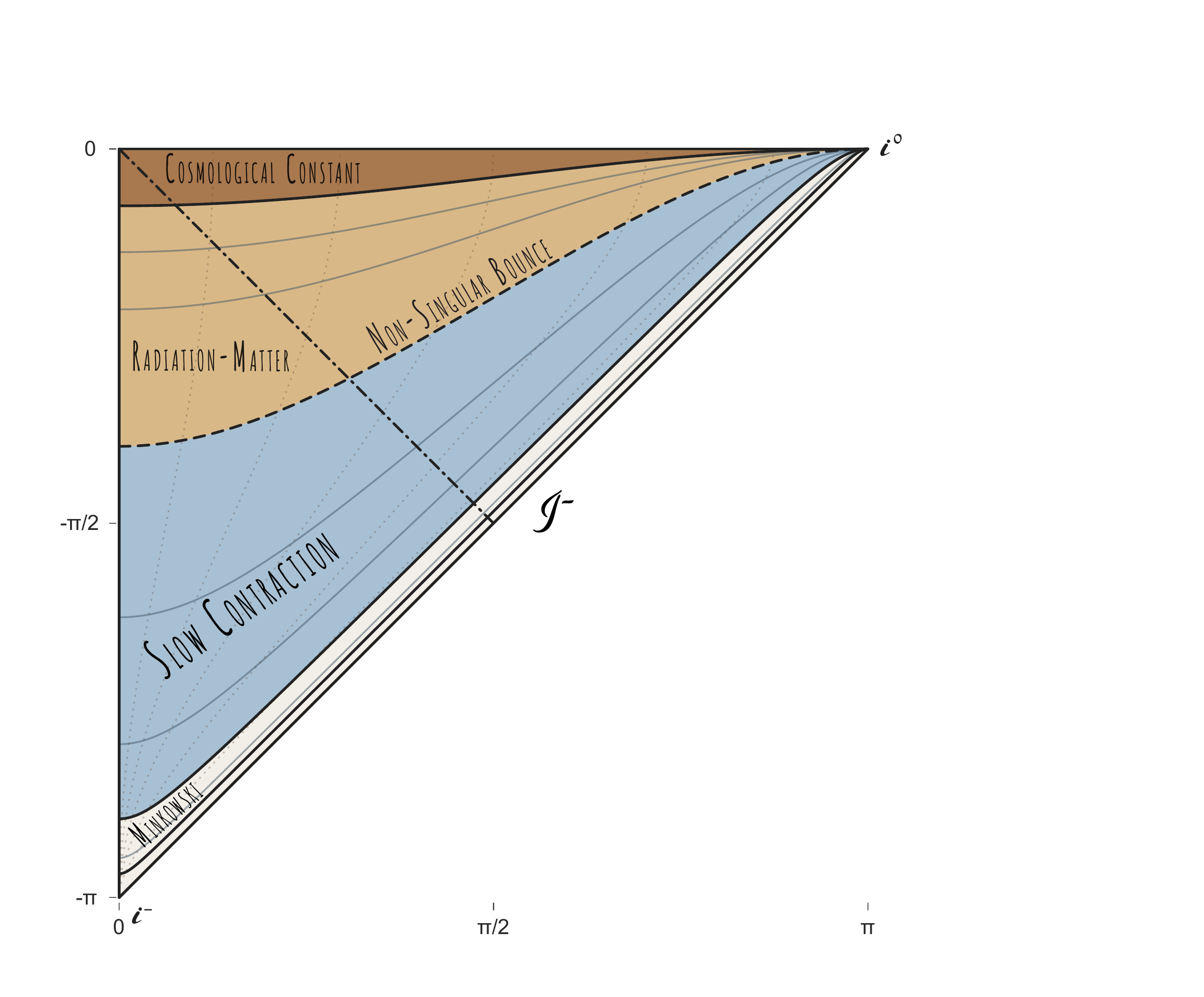}
        \caption{Figure 2. Conformal Diagram  in $\eta-\chi$ plane  of a universe with an asymptotic past Minkowski, a slow contraction phase ending with a non singular bounce, and smoothly transitioning into  radiation-matter domination followed by dark energy  (cosmological constant) domination in the asymptotic future. The event horizon in this case is indicated by the diagonal dot-dashed line that intersects the upper left-hand corner. }
        \label{fig:second}
    \end{subfigure}

    \caption*{}
    \label{fig:both}
\end{figure}

\section{Effect of an additional scalar field: stability versus instability}
\label{sec:7}

So far, our discussion of causal structure has focused on the evolution of the contracting cosmological background assuming a single stress-energy component, a scalar field that drives slow contraction, or a cosmological constant for the  contracting de Sitter case.
However, slow contraction typically is described by two or more scalar fields, one of which drives the background evolution and the other(s) of which are responsible for generating growing or constant-mode fluctuations of quantum origin responsible for the density fluctuations observed in the cosmic microwave background \cite{Buchbinder:2007ad,Lehners:2008vx,DiMarco:2002eb,Levy:2015awa,Ijjas:2021ewd}, though two fields are not strictly required: a single field can suffice if the spectrum is instead generated through a different mechanism, such as a string-motivated S-brane transition \cite{Brandenberger:2020xww}.  
 
In this section, we consider the effect of adding a canonical scalar
field whose attractor solution is characterized by an equation-of-state
parameter $\varepsilon'$.  Depending on the scalar-field potential,
$\varepsilon'$ can take an arbitrary positive value. We compare how adding such
a field affects the dynamical stability of a slow-contraction phase versus a
contracting de~Sitter phase.
 
\paragraph{Slow contraction}

The inclusion of additional scalar degrees of freedom does not spoil the
slow-contraction attractor.  Suppose, for example, that a second scalar field is
added whose would-be attractor solution is characterized by equation-of-state
parameter $\varepsilon'$.  If $\varepsilon'<\varepsilon$, then the added field
does not overtake the original slow-contracting component; if it starts
subdominant, it remains subdominant, and the attractor solution is the same as
without the added field.  If $\varepsilon'=\varepsilon$, the relative
contribution of the added field is marginal and the solution remains in the
same slow-contraction class.  If instead $\varepsilon'>\varepsilon$, the added
field can eventually come to dominate.  But this does not destabilize the
solution: the background is then driven toward a new slow-contracting attractor
with larger $\varepsilon'$.  Since larger $\varepsilon'$ corresponds to a more
slowly contracting scale factor and a faster growth of the dominant energy
density relative to curvature, anisotropy, and inhomogeneity, the new attractor
smooths and flattens even more efficiently.

Thus, additional scalar fields do not destabilize the slow-contracting phase.
They either remain subdominant, leave the original slow-contraction class
unchanged, or replace it by an even stronger slow-contraction attractor.

This background-level argument has been substantially \\ strengthened beyond linear order. Non-perturbative numerical relativity simulations confirm that a sufficiently steep slow-contraction potential drives the Einstein-scalar system to an ultralocal state in which spatial gradients become dynamically negligible at each point in space, after which the geometry converges to a homogeneous, flat, isotropic attractor even when starting from large, non-perturbative anisotropy, spatial curvature, and inhomogeneity \cite{Belinski:1970,Ijjas:2021uxr}. This robustness extends to the multi-field case relevant for generating the observed density spectrum: simulations of slow contraction driven by two kinetically-coupled scalar fields -- the type required to produce a nearly scale-invariant curvature spectrum consistent with CMB observations -- show that the same flat, homogeneous, isotropic attractor is reached for a wide range of inhomogeneous and anisotropic initial conditions \cite{Ijjas:2021tul}, and this stability persists \\ through the bounce, where the entropic perturbations generated during slow contraction source the curvature spectrum and become negligible by the end of the transition \cite{Ijjas:2021ewd,Ijjas:2020xtn}. (Although Tolley and Wesley have pointed out examples in which an added second field can produce unstable growth of isocurvature modes \cite{Tolley:2007nq}, such models are not in the class of models with a semi-infinite slow-contraction phase considered here).

\paragraph{Contracting de~Sitter}

By contrast, contracting de~Sitter space exhibits the opposite behavior.
The background vacuum energy density is given by the cosmological constant,
$\varrho_{\Lambda}=\text{const}$, corresponding to $\varepsilon=0$.  Any
canonical scalar field with non-vanishing kinetic energy has
$\varepsilon'>0$, so in a contracting phase its contribution grows relative to
the constant vacuum energy.  Hence, even an initially subdominant scalar field
generically comes to dominate, driving the spacetime away from de~Sitter.

Thus, while contracting de~Sitter space may be geodesically complete as a
pure background solution, it is dynamically unstable once an extra scalar field, anisotropy, or inhomogeneities are included.  
This instability is not a slicing artifact but an intrinsic dynamical property of
a contracting de Sitter phase, which is in sharp contrast with the stable attractor behavior of slow contraction.

\section{Geodesic completeness versus particle horizons:
slow contraction and global de Sitter}
\label{sec:8}

It is important to distinguish between past geodesic completeness and
the existence of a particle horizon. These are logically
independent properties, and their distinction reveals a sharp and
physically significant contrast between slow contraction and
global de Sitter space.

A spacetime is {\it past geodesically complete} if every past-directed
timelike and null geodesic has infinite proper time or affine parameter.
The existence of a particle horizon at a time $t_0$ is determined
by the comoving distance
\begin{equation}
\chi_P(t_0) = \int_{t_i}^{t_0} \frac{dt}{a(t)} ,
\label{eq:PH}
\end{equation}
where $t_i$ denotes some initial time.
If this integral converges, the particle horizon is finite; conversely, 
there is no particle horizon if Eq.\eqref{eq:PH} diverges.

\paragraph{Slow contraction ($\varepsilon >3$)}
As shown in Section~\ref{sec:3}, the particle-horizon integral diverges
when the slow-contraction phase extends to $t\rightarrow -\infty$.
Consequently, arbitrarily large comoving regions can be in causal contact  in
the past, independently of the spatial curvature of the FLRW slicing, provided
the conformal-time integral diverges.

The question of past geodesic completeness is distinct.
In spatially flat or open slow-contraction solutions sourced by a canonical
scalar field with an appropriate potential producing a semi-infinite period of contraction, the spacetime approaches
Minkowski space asymptotically as $t\rightarrow -\infty$.
In this case, all curvature invariants vanish, and both timelike and null
geodesics have infinite proper time or affine parameter.
As a result, slow contraction in flat or open geometries can be {\it past geodesically
complete} while simultaneously possessing no particle horizon.

The discussion above concerns the exactly homogeneous and isotropic (zero-shear) slow-contraction solution. If homogeneous anisotropy (shear) is present, its evolution depends on spatial curvature. In open universes, spatial curvature ($\varrho_k\propto a^{-2}$) dominates over both the slow-contraction field ($\varrho_\phi\propto a^{-2\varepsilon}$, $\varepsilon>3$) and homogeneous shear ($\varrho_\sigma\propto a^{-6}$) as $a\rightarrow\infty$, so the isotropic Minkowski limit is approached for a wide basin of initial shear values. In spatially flat universes, homogeneous shear is not suppressed under backward evolution, and a generic solution instead approaches a vacuum Kasner solution as $a\rightarrow\infty$. This does not affect geodesic completeness, however: Kasner-type solutions have curvature invariants that vanish as $a\rightarrow\infty$, so past-directed geodesics retain infinite affine length even in this anisotropic case. Thus, past geodesic completeness holds generically in flat or open geometries, while approach to the isotropic Minkowski/Bunch-Davies vacuum specifically requires open curvature or exactly zero homogeneous shear.

\medskip

\noindent
{\bf Remarks.}
Our conclusions do not require that the scalar field which drives slow contraction  dominates all
the way to $t\rightarrow-\infty$.
If matter or negative spatial curvature becomes dominant sufficiently far in the
past so that the solution asymptotes to a Minkowski limit with $|H|\rightarrow 0$ and
vanishing curvature invariants, then the spacetime remains past geodesically
complete. 
At the same time, no particle horizon emerges:
whenever the past asymptotics satisfy $a(t)\sim |t|^{1/\varepsilon}$ with $\varepsilon>3$, the conformal-time
integral $\int_{-\infty}^{t_0} dt/a(t)$ diverges, and the cosmology is
horizon-free.

By contrast, in closed slow-contracting solutions, the energy density driving slow contraction eventually becomes subdominant to positive spatial curvature in the
far past.
In such cases, the spacetime typically encounters a curvature-dominated
singularity at finite proper time, rendering it past geodesically incomplete,
even though a particle horizon may not be present up to that point.
This illustrates that geodesic completeness in slow contraction depends on the
global evolution and spatial curvature, while the absence of a particle-horizon 
depends only on the behavior of the scale factor which determines the conformal-time integral associated with the particle horizon.

\paragraph{Global de Sitter}
 Expressed in global coordinates, the scale factor of closed de~Sitter space is given by 
{$a(t)=H_\Lambda^{-1}\cosh(H_\Lambda t)$}. This leads to a past geodesically complete spacetime: all past-directed timelike and null geo\-desics have infinite proper time or affine length.
Nevertheless, as shown in Section~\ref{sec:5}, the particle-horizon integral
from $t=-\infty$ to the neck at $t=0$ is finite.
Consequently, even after a semi-infinite period of contraction, only a finite
portion of the spatial hypersurface $S^3$ lies within any observer’s past light cone.

This example demonstrates that geodesic completeness does not imply 
the absence of a particle horizon.
Global de~Sitter space is past complete yet possesses a finite particle horizon,
whereas slow contraction in flat or open geometries can be both past complete
and horizon-free.

\section{Evading the Borde--Guth--Vilenkin theorem}
\label{sec:9}

A semi-infinite slow contraction phase  evades  the
Borde--Guth--Vilenkin (BGV) theorem~\cite{Borde:2001}, the standard no-go
result invoked against past-eternal cosmologies, and does so while remaining
both past geodesically complete (Section~\ref{sec:8}) and dynamically
stable (Section~\ref{sec:7}).

BGV is a purely kinematic result: it assumes neither the Einstein equations nor
any energy condition, so the null-energy-condition violation required for a
nonsingular bounce is irrelevant to whether it applies. 

Let $\mathcal{O}$ be a
past-directed null or timelike geodesic, with affine parameter (null case) or
proper time $\tau$ (timelike case), and let $u^\mu$ be the four-velocity of a
comoving congruence defined along $\mathcal{O}$. We orient $\tau$ so that $\tau_f$ corresponds to a fixed reference event ({\it e.g.}, \ a present-day observer) and $\tau_i < \tau_f$ to an earlier event along $\mathcal{O}$; the integral below is taken in the conventional forward sense, from $\tau_i$ to $\tau_f$, while the geodesic itself is extended past-ward by sending $\tau_i \to -\infty$. Defining
$\gamma \equiv u_\nu v^\nu$ (with $v^\mu = dx^\mu/d\tau$ the tangent to
$\mathcal{O}$), the Hubble parameter measured along $\mathcal{O}$ can be written
as a total derivative, ${\cal H} \equiv dF/d\tau$, with
\begin{equation}
F(\gamma) =
\begin{cases}
\gamma^{-1}, & \text{null },\\[4pt]
\tfrac{1}{2}\ln\! \left(\dfrac{\gamma+1}{\gamma-1}\right), & \text{timelike }.
\end{cases}
\label{eq:bgvF}
\end{equation}
Integrating along $\mathcal{O}$ from $\tau_i$ to  $\tau_f$,
\begin{equation}
\int_{\tau_i}^{\tau_f} {\cal H}\, d\tau = F(\gamma_f) - F(\gamma_i) \le F(\gamma_f).
\label{eq:bgvbound}
\end{equation}
Defining the averaged expansion rate
${\cal H}_{\rm av} \equiv (\tau_f-\tau_i)^{-1}\!\int {\cal H}\,d\tau$, the theorem states that
if ${\cal H}_{\rm av}>0$ along a past-directed null or noncomoving timelike
geodesic, then $\int {\cal H}\,d\tau$ is bounded and the geodesic has finite affine or
proper length, {\it i.e.},\ it is past-incomplete.

 Eq.~\eqref{eq:bgvbound} is based on the
behavior of $\gamma$, the relative Lorentz factor (timelike) or redshift factor
(null) between $\mathcal{O}$ and the comoving frame.  Along a complete past-directed
geodesic, the affine parameter or proper time $\tau \to -\infty$, and the corresponding
cosmological time $t$ at events on $\mathcal{O}$ also tends to $-\infty$.
Since the peculiar
momentum of a noncomoving geodesic scales as $\pi\propto 1/a(t)$, the past
behavior of $\gamma$ is governed entirely by the scale factor along the geodesic. In the
inflationary or contracting-de~Sitter case that BGV was designed to address,
$a\to0$ toward the past, so $\pi$ and hence $\gamma$ blueshift without
bound; $F(\gamma_i)\to 0$ (null) and the integral in
Eq.~\eqref{eq:bgvbound} is bounded, forcing past-incompleteness.

Slow contraction behaves the opposite way. From the background
solution $a(t)=a_0(-t)^{1/\varepsilon}$ of Section~\ref{sec:2}, the scale
factor grows without bound
toward the far past, $a(t)\to\infty$ as $t\to-\infty$, and therefore also along
the geodesic as $\tau\to-\infty$. This means, a noncomoving geodesic
 redshifts toward the past, $\pi\to0$ and $\gamma\to1$. In the
timelike case $F(\gamma)$ diverges as $\gamma\to1$, so the right-hand side of
Eq.~\eqref{eq:bgvbound} does not bound the integral; in the null case the
analogous boundary term likewise fails to produce a finite affine length. The
BGV bound simply does not constrain these geodesics, in full consistency with
the explicit past completeness established in Section~\ref{sec:8}.
Equivalently, since the phase contracts ($H = \dot a/a = 1/(\varepsilon t)<0$ for $t<0$)
and approaches the Minkowski limit $H\to0^-$ as $t\rightarrow -\infty$, the averaged expansion rate along
past-directed geodesics is non-positive, ${\cal H}_{\rm av}\le 0$, and the underlying assumption of the theorem is not met.
Note that our conclusion does not depend on the spatial-curvature or tuning
assumptions discussed in Sections~\ref{sec:7} and~\ref{sec:8}.
Whether the far past is governed by the  scalar field driving slow contraction, by pressureless
matter with $\varepsilon=3/2$, or by negative spatial curvature with effective $\varepsilon =1$, the scale
factor grows as $t\to-\infty$ and  $a\to\infty$, so $\gamma\to1$ and the
evasion holds in every case. This includes both an open or matter-dominated far past and the spatially-flat case of Section~\ref{sec:8}.

The same argument applies if the bounce is followed by an arbitrarily large but finite number of additional cycles before reaching a present-day expanding phase: each cycle of finite duration contributes a bounded amount to both the affine length and the integral $\int {\cal H}\,d\tau$, so the past-averaged expansion is determined by the semi-infinite slow-contraction tail and remains non-positive, while $\gamma\to1$ at the past end of the geodesic regardless of the intervening cyclic episode. Finiteness of the number of cycles is essential: BGV's original analysis of cyclic models~\cite{Borde:2001} concerns the opposite limit of infinitely many cycles with net expansion per cycle, in which the positive contributions accumulate without bound and the theorem applies.

\section{Conclusions}
\label{sec:10}
We have analyzed the causal and global structure of contracting cosmologies,
with particular emphasis on slow contraction and its contrast with contracting
de~Sitter space.
Our primary technical result is that a sufficiently long phase of slow
contraction with equation of state $\varepsilon > 3 $  eliminates the particle horizon: the
comoving particle-horizon integral diverges, demonstrating that any two spatial points with finite comoving separation can have overlapping past light cones at sufficiently
early times.
A central conceptual lesson of this work is the distinction between particle
horizons and geodesic completeness.
Slow contraction in spatially flat or open geometries can be past geo\-desically complete and asymptote to Minkowski spacetime in the far past, while simultaneously
possessing no particle horizon.
This conclusion does not require that the stress-energy driving slow contraction dominate the total energy density all the way to
$t\to -\infty$. If matter or negative  spatial curvature (open geometry) governs the
asymptotic past with effective equation of state $\varepsilon \ge 1$, the spacetime still
approaches a Minkowski limit, remains past geodesically complete, and the
conformal-time integral diverges. The cosmology is therefore free of a particle horizon.
The contracting phase of global de~Sitter space provides a sharp contrast: despite being past
geodesically complete in closed slicing, it retains a finite particle horizon,
so that only a limited region is ever causally connected.
This demonstrates explicitly that geodesic completeness alone does not determine
the causal structure of a cosmological phase.
Beyond avoiding the horizon problem, slow contraction exhibits a unique combination of properties that distinguish it from  other cosmological settings:
\begin{enumerate}
    \item[(1)] the particle horizon is absent;
    \item[(2)] spacetime is past geodesically complete in spatially flat or open geometries;
    \item[(3)] evolution towards the bounce is driven robustly towards an attractor solution; the analogous backward evolution towards the asymptotic past is likewise robust in open geometries, where spatial curvature dominates over homogeneous anisotropy, whereas in flat geometries the isotropic limit requires exactly homogeneous and isotropic initial data (see Sec.~\ref{sec:8}) \cite{EricksonWesleySteinhardtTurok2004}; 
    \item[(4)] in open geometries, or in flat geometries with exactly homogeneous and isotropic initial data, the geometry approaches Minkowski spacetime and a natural Bunch-Davies-like vacuum spectrum of quantum fluctuations;
    \item[(5)] the semi-infinite phase of slow contraction evades the Borde--Guth--Vilenkin theorem;
    \item[(6)] the homogeneous background is perturbatively stable in the presence of additional fields;
    \item[(7)] the gravitational sector remains in a low-entropy state throughout slow contraction \cite{Penrose1989NYAS,PenroseRoadToReality2004,IjjasSteinhardt2021Entropy}.
\end{enumerate}
\noindent
A general take-away  is that a standard Big Bang cosmology that is  past geodesically incomplete and generically produces initial conditions with high gravitational entropy, anisotropy, inhomogeneity, and spatial curvature  can be converted into a cosmology that is past geodesically complete in flat or open geometries, with a smooth, flat, Weyl-curvature-suppressed (gravitationally low entropy) start to the expanding phase, by replacing the bang with a semi-infinite slow-contraction phase and a non-singular bounce. To our knowledge, no other known cosmology, including a singular Big Bang (with or without inflation) or a contracting de~Sitter phase, exhibits this combination of causal, geometric, dynamical, quantum-mechanical, and semiclassical consistency properties.

\section*{Acknowledgments}
\noindent
We thank V. Mukhanov for useful discussions. This work is supported in part by the U.S. Department of Energy under grant number
DE-FG02-91ER40671, by the Simons Foundation under grant number 654561, and by the John Templeton Foundation under grant number 63749. The opinions expressed in
this publication are those of the author(s) and do not necessarily reflect the views of the
John Templeton Foundation.  

\onecolumn
\appendix
\section{Comparison: slow contraction vs.\ contracting de Sitter with multiple scalar fields}

\noindent
Semi-infinite slow contraction \textit{with a scalar field plus potential} is past complete,
horizon-free, and dynamically smoothing, whereas semi-infinite contracting de~Sitter
\textit{with a scalar field plus potential} is horizon-limited, dynamically unstable,
and generically crunches.
Here and throughout this table, the ``background'' refers to the homogeneous,
isotropic FLRW solution with zero spatial curvature, obtained in the absence of
any additional scalar fields (except where explicitly indicated) aside from the one responsible for smoothing.

\vspace{0.4em}

{\fontsize{9}{11}\selectfont
\begin{adjustbox}{width=\textwidth}
\begin{tabularx}{\textwidth}{L Y Y}
\toprule
\textbf{Feature} &
\shortstack{\textbf{Contracting de~Sitter }\\\textbf{background+ scalar field}} &
\shortstack{\textbf{Slow-contraction}\\\textbf{background + scalar field}} \\
\midrule

Background equation of state
& $\varepsilon=0$ (vacuum energy)
& $\varepsilon>3$ (ultra-stiff slow contraction component) \\

Background scale factor
& $a(t)\propto e^{-H_\Lambda t}$
& $a(t)\propto(-t)^{1/\varepsilon},\;\varepsilon>3$ \\

Background curvature scale in far past
& Finite ($|R|\sim H_\Lambda^2$), constant everywhere
& Vanishes ($R\rightarrow0$; Minkowski limit) \\

Past geodesic completeness (with scalar field)
& Yes (scalar energy redshifts away as $t\rightarrow-\infty$)
& Yes (asymptotes to Minkowski space) \\

Stable evolution when other scalar fields or matter are added
& No; scalar kinetic energy or matter energy comes to dominate and induces a crunch
& Smooth approach to bounce \\

\addlinespace[4pt]
No causal horizon?
& False (conformal time integral converges)
& True (conformal time integral diverges) \\

Anisotropy (shear)
& $\varrho_\sigma\propto a^{-6}$ unsuppressed
& $\varrho_\sigma/\varrho_\phi\rightarrow0$; anisotropy dynamically damped \\

No-hair–type attractor?
& No; growing scalar field kinetic energy and shear modes
& Yes; flat, homogeneous, isotropic attractor \\

Initial-condition sensitivity
& Sensitive to initial conditions
& Wide basin of attraction \\

Linear perturbative stability
& Generically unstable to scalar and metric perturbations
& Stable adiabatic background; entropy modes controllable \\

Vacuum selection
& No dynamically selected preferred vacuum 
& Minkowski past with Bunch-Davies-like vacuum on all wavelengths \\

Generic outcome at the end of contraction
& Departure from de~Sitter and evolution toward a crunch
& Smooth, flat, isotropic state reached at the bounce \\

Gravitational entropy at the end of contraction
& Large; anisotropy and gradient growth generate high entropy
& Negligible;  smoothing by slow contraction yields a highly ordered state \\

\bottomrule
\end{tabularx}
\end{adjustbox}
}

\newpage

\bibliographystyle{elsarticle-num}
\bibliography{refs}

\end{document}